\documentclass[aps,prc,twocolumn,showpacs,superscriptaddress,nofootinbib]{revtex4-2}
\usepackage{amsmath,amssymb,graphicx}
\usepackage{url}
\usepackage{slashed}
\usepackage{bm}
\newcommand{\nopieft}{\mbox{$\slashed{\pi}$EFT}}

\newcommand{\be}{\begin{equation}}
\newcommand{\ee}{\end{equation}}

\newcommand{\fm}{\,\text{fm}}
\newcommand{\ifm}{\,\text{fm}^{-1}}

\newcommand{\etal}{\emph{et al.\,}}

\begin{document}

\title{Few-Nucleon Systems within Finite-Cutoff Pionless EFT}

\author{Liron H. Avraham}
\affiliation{The Racah Institute of Physics, The Hebrew University, Jerusalem 9190401, Israel}

\author{Betzalel Bazak}
\affiliation{The Racah Institute of Physics, The Hebrew University, Jerusalem 9190401, Israel}

\date{\today}

\begin{abstract}
We investigate pionless effective field theory (\nopieft) with finite-cutoff regularization as a framework for describing few-nucleon systems. This formulation incorporates effective-range effects already at leading order (LO), thereby reaching next-to-leading-order (NLO) accuracy while maintaining computational efficiency. Using correlated-Gaussian stochastic variational methods in a weak harmonic-oscillator trap, together with neutral and Coulomb-modified quantization conditions, we calculate binding energies and low-energy $S$-wave scattering parameters for systems with up to five nucleons. At an optimal cutoff, the computed binding energies of the deuteron, triton, helion, and alpha particle reproduce experimental values at the percent level once a three-body force is included. Scattering parameters for proton--proton, nucleon--deuteron, nucleon--triton, proton--helion, deuteron--deuteron, and nucleon--alpha channels are obtained and found to be consistent with both experimental data and existing NLO \nopieft\ calculations. These results demonstrate that finite-cutoff \nopieft\ offers a robust and predictive framework for few-body nuclear physics.
\end{abstract}

\maketitle

\section{Introduction}
Effective field theories (EFTs) provide a systematic and model-independent framework for describing low-energy phenomena in terms of the relevant degrees of freedom, while encoding unresolved short-distance physics into contact interactions \cite{Wei90}. 
In nuclear physics, the pionless EFT (\nopieft) is particularly well suited for few-nucleon systems at momenta well below the pion mass \cite{KapSavWis98,BedHamKol99}. 
Its dynamics are governed by contact operators and their derivatives, organized according to a well-established power-counting scheme. 
A comprehensive review of EFT applications in nuclear physics is given in Ref.~\cite{HamKonKol20}.  

Contact interactions require regularization, typically implemented with a momentum cutoff $\Lambda$. 
Predictivity is ensured once observables become independent of $\Lambda$, which defines the renormalized limit of the theory. 
Although the formal prescription is $\Lambda \to \infty$, in practice---especially in numerical many-body calculations---finite values of $\Lambda$ are routinely employed \cite{HamKonKol20}. 

As an illustrative example, consider the deuteron in \nopieft.  
At leading order (LO), the two-body potential in the spin–triplet channel is represented by a smeared delta function of width $\Lambda^{-1}$, with its low-energy constant (LEC) fitted to a two-body observable such as the $n$--$p$ scattering length.  
Both the interaction range and the effective range then scale as $\Lambda^{-1}$, recovering the zero-range limit as $\Lambda \to \infty$.  
At next-to-leading order (NLO), operators with derivatives are included, with LECs tuned to reproduce the empirical effective range.  
Notably, for certain finite values of $\Lambda$, the empirical effective range is already reproduced at LO, rendering the explicit NLO correction effectively redundant.

Figure~\ref{fig:d} illustrates this for the deuteron ground-state energy, where LO and NLO results converge with the experimental value of $-2.225$~MeV \cite{HuaWanKon21} at $\Lambda \approx 1.25 \ifm$.  

\begin{figure}
    \centering
    \includegraphics[width=\columnwidth]{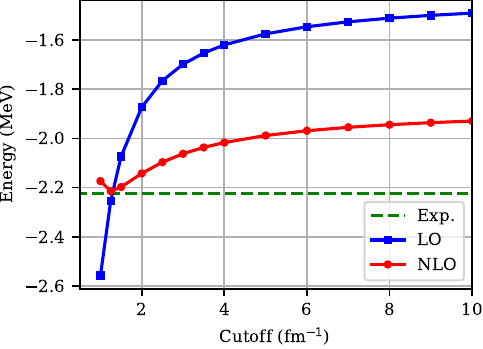}
    \caption{Deuteron ground-state energy in \nopieft\ as a function of the cutoff $\Lambda$. 
    The LO (blue squares) and NLO (red dots) results coincide with the experimental value (green dashed line) at $\Lambda \approx 1.25 \ifm$, where the empirical effective range is already reproduced at LO.}
    \label{fig:d}
\end{figure}

This observation raises the question of whether calculations performed directly at such a cutoff---where LO interactions already reproduce NLO behavior---can provide NLO-quality results with reduced numerical effort.  

A related strategy was recently proposed in Ref.~\cite{ConSchGne25}, where an artificial range is introduced at LO and subsequently removed perturbatively at NLO, yielding a stable foundation for higher-order improvements. 
In practice, this formulation is equivalent at LO to calculations performed at a specific finite cutoff, as also explored in Refs.~\cite{KirGriShu10,KieGat16,KieVivLog18,SchGirGne21,GatKie23}.  

Traditionally, calculations at multiple cutoffs are performed to test renormalization-group (RG) invariance and thereby validate the power counting of the theory. 
Such studies have established that the three-body force must be promoted to LO \cite{KapSavWis98,BedHamKol99}, while a four-body force first appears at NLO \cite{BazKirKon19}, contrary to naive dimensional analysis. 
Once the power counting is verified, however, accurate and predictive results can be obtained at a single, well-chosen cutoff.  

Another benefit of varying the cutoff is to assess the residual cutoff dependence. Since this dependence is expected to be removed by higher-order terms, its variation can be taken as a lower bound on the theoretical uncertainty due to the omitted contributions \cite{Har20}. 
Our method lacks this feature and therefore must rely on more traditional approaches to estimate the theoretical uncertainty, such as evaluating the expansion parameter, which suggests an uncertainty of order 10\% for \nopieft\ at NLO.

In this work, we apply finite-cutoff \nopieft\ to few-nucleon systems with up to five nucleons, computing binding energies and low-energy $S$-wave scattering parameters.  
We demonstrate that this strategy reproduces NLO \nopieft\ results and experimental data, indicating that it provides a practical framework for few-body nuclear physics.

\section{Model and Methods}
We employ the LO \nopieft\ interaction regulated at a finite cutoff, consisting of two- and three-body potentials,
\begin{equation} 
V = \sum_{i<j} V_{2B}(r_{ij}) + \sum_{i<j<k} \sum_{\text{cyc}} V_{3B}(r_{ij},r_{jk})\,.
\end{equation}

The two-body potential is given by
\begin{equation}
    V_{\text{2B}}(r_{ij}) = 
      C_s \,\delta_{\Lambda_s}(r_{ij}) \,\hat{\mathcal{P}}^s_{ij}
    + C_t \,\delta_{\Lambda_t}(r_{ij}) \,\hat{\mathcal{P}}^t_{ij}
    + \frac{e_i e_j}{r_{ij}}\,,
\end{equation}
where $\hat{\mathcal{P}}^s_{ij}$ and $\hat{\mathcal{P}}^t_{ij}$ are spin--isospin projectors onto the singlet ($S=0$, $I=1$) and triplet ($S=1$, $I=0$) two-nucleon channels, respectively, and $\delta_{\Lambda}(r)=\exp(-\Lambda^2r^2/4)$ denotes a non-normalized Gaussian-regulated delta function with cutoff $\Lambda$. 

The static Coulomb interaction acts only between charged particles, with $e_i = 0$ for neutrons and $e^2 = \alpha \hbar c = 1.44$~MeV~fm for a proton pair. Formally, a dedicated $p$–$p$ term must be included to properly renormalize $S$-wave $p$--$p$ scattering \cite{KonRav00,VanEgoKer14,RojSch25}. However, at the chosen finite cutoff this LEC plays no significant role, as we demonstrate later that the $p$--$p$ scattering parameters are well reproduced without it.

The three-body potential takes the form
\begin{equation}
    V_{\text{3B}}(r_{ij},r_{jk}) = 
    D\,\delta_{\Lambda_3}(r_{ij})\,\delta_{\Lambda_3}(r_{jk})\,\hat{\mathcal{P}}^{1/2}_{ijk}\,,
    \label{eq:3b}
\end{equation}
where $\hat{\mathcal{P}}^{1/2}_{ijk}$ projects onto the spin-$1/2$, isospin-$1/2$ three-nucleon channel.

The two-body LECs $C_s$, $C_t$ and cutoffs $\Lambda_s$, $\Lambda_t$ are fitted to reproduce the $S$-wave scattering lengths ($a_s = -18.95 \fm$, $a_t = 5.419 \fm$) and effective ranges ($r_s = 2.75 \fm$, $r_t = 1.753 \fm$) \cite{Hac06}. 
The three-body LEC $D$ and cutoff $\Lambda_3$ are adjusted to match the binding energies of the triton ($B(^3\text{H}) = 8.482$ MeV), helion ($B(^3\text{He}) = 7.718$ MeV), and alpha particle ($B(^4\text{He}) = 28.296$ MeV) \cite{HuaWanKon21}. 

The resulting LECs and cutoff values are summarized in Table~\ref{tab:LECs_Lambdas}. Different cutoffs were chosen for each channel; however, since all cutoffs lie within the validity regime of \nopieft, the results remain within the theoretical uncertainty of the framework.

\begin{table}
  \caption{\label{tab:LECs_Lambdas} 
  Calculated LECs (in MeV) and finite cutoffs (in $\ifm$) employed in our model. The normalization of the Gaussian-regulated delta function is absorbed into the LECs.}
  \begin{ruledtabular}
  \begin{tabular}{ccccc}
    LECs (MeV)       & $C_s$       & $C_t$       & $D$         \\ \hline
                     & $-31.246$   & $-67.599$   & $16.4$      \\ \hline\hline
    Cutoffs ($\ifm$) & $\Lambda_s$ & $\Lambda_t$ & $\Lambda_3$ \\ \hline
                     & $1.11479$   & $1.29624$   & $1.78885$   \\
  \end{tabular}
  \end{ruledtabular}
\end{table}

We solve the few-body Schrödinger equation using the stochastic variational method (SVM) with correlated Gaussian basis functions \cite{SuzVar98,BazEliKol16}, which efficiently capture both short-range correlations and the long-range behavior of weakly bound states.

Since SVM, like other variational approaches, is most effective for compact bound states, we embed the systems in a weak external harmonic-oscillator (HO) trap of the form
\begin{equation}
    V_{\rm HO}(\mathbf{r}) = \frac{m\omega^2}{2A} \sum_{i<j} (\mathbf{r}_i-\mathbf{r}_j)^2\,,
\end{equation}
with oscillator frequency $\omega$.  

We consider the scattering of two bound subclusters $B$ and $C$ inside the trap. 
Choosing the trap length $\sqrt{2\hbar/(m\omega)}$ much larger than all intrinsic length scales ensures that the subclusters behave effectively as pointlike particles. 
The trapped wave function is then matched to the analytic solution for two trapped particles with short-range interactions \cite{BusEngRza98}. 

For the case of neutral particles, the $S$-wave phase shift $\delta$ at relative momentum $k$ is extracted from
\begin{equation}
    k \cot \delta = -\sqrt{4\mu \omega/\hbar}\, 
    \frac{\Gamma\!\left[(3-2\epsilon)/4\right]}{\Gamma\!\left[(1-2\epsilon)/4\right]}\,,
    \label{Busch}
\end{equation}
where $\mu = m_B m_C / (m_B + m_C)$ is the reduced mass of the clusters, 
$\Gamma(x)$ is the Gamma function, 
$k = \sqrt{2\mu\hbar\omega\epsilon}$ is the relative momentum, and 
$\epsilon = (E_A - E_B - E_C)/(\hbar\omega)$ is the dimensionless energy of the trapped $A$-body system relative to the $B+C$ threshold. 
Bound-state energies $E_A$, $E_B$, and $E_C$ are computed with SVM.  

The scattering length $a$ and effective range $r$ are obtained by fitting the extracted phase shifts to the effective-range expansion (ERE),
\begin{equation} \label{eq:ere}
k \cot \delta = -\frac{1}{a} + \frac{1}{2} r \, k^2 + \mathcal{O}(k^4)\,.
\end{equation}

This formalism was employed in Ref.~\cite{SchBaz23} to determine neutral few-nucleon ($A \leq 4$) scattering parameters within NLO \nopieft, and later extended to $n$--$\alpha$ $S$-wave scattering in Ref.~\cite{BagSchBaz23}.  

Extending the method to charged clusters requires accounting for the long-range Coulomb interaction. 
In this case, the relation between trapped spectra and free-space phase shifts [Eq.~\eqref{Busch}] must be replaced by its Coulomb-modified counterpart, and the effective-range expansion [Eq.~\eqref{eq:ere}] must be generalized to include Coulomb effects.

The Coulomb-modified effective-range expansion reads \cite{Bet49}
\begin{align}\label{eq:coulomb_ere}
    \kappa(\eta) &\equiv 
    C_0^2(\eta)\,k\,\cot \delta + 2k\eta\,h(\eta) \nonumber \\
    &= -\frac{1}{a} + \frac{1}{2} r\,k^2 + \mathcal{O}(k^4),
\end{align}
where $C_0^2(\eta) = 2\pi\eta/(e^{2\pi\eta}-1)$, $h(\eta) = \text{Re}[\Psi(1+i\eta)] - \ln(\eta)$ with $\Psi$ the digamma function, and $\eta = Z_1 Z_2 \mu e^2 /(\hbar^2 k)$ is the Sommerfeld parameter where $Z_i$ is the number of protons in cluster $i$.

The $S$-wave phase shifts of two charged clusters can then be obtained from the trapped spectrum using the Coulomb-modified quantization condition \cite{Guo21}:
\begin{align}\label{bush_c}
    &-2\mu\,C_0^2(\eta)\,k \cot \delta = \nonumber \\
    &\qquad \lim_{r,r'\to 0} \Big\{
        \text{Re}\big[G^{C,\infty}_0(r,r';\epsilon)\big] 
        - G^{C,\omega}_0(r,r';\epsilon)
      \Big\},
\end{align}
where $G^{C,\omega}_0$ and $G^{C,\infty}_0$ denote the Coulomb Green’s functions in the trap and in free space, respectively.  

Accurate formulations and numerical implementations of the Coulomb-modified quantization condition have been developed in Refs.~\cite{Guo21,BagBarRoj25}.  
Most recently, this framework was applied within NLO \nopieft\ to extract low-energy scattering phase shifts in the $p$--$d$, $d$--$d$, and $p$--$h$ systems \cite{RojSch25}.  
In our work, we first compute the few-body spectra including the Coulomb interaction using SVM, and then extract the corresponding scattering parameters following the implementation of Ref.~\cite{BagBarRoj25}.

As a representative example, Fig.~\ref{fig:Nd32} shows the effective-range expansion obtained for the neutral $n$--$d$ and charged $p$--$d$ $S=3/2$ scattering channels, using Eqs.~\eqref{eq:ere} and \eqref{eq:coulomb_ere}, respectively.  
In both cases, the effective-range expansion is well satisfied, enabling a reliable extraction of the scattering parameters. 

\begin{figure}
    \centering
    \includegraphics[width=\columnwidth]{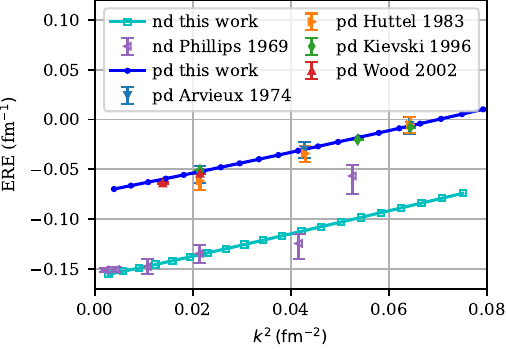}
    \caption{Effective-range expansion for nucleon--deuteron scattering in the $S=3/2$ channel. 
    Results for neutral $n$--$d$ scattering (cyan squares, Eq.~\eqref{eq:ere}) and charged $p$--$d$ scattering (blue dots, Eq.~\eqref{eq:coulomb_ere}) are shown together with second-order polynomial fits (solid lines). 
    Available experimental data are also included: Phillips \etal~\cite{PhiBar69} for $n$--$d$, and Arvieux~\cite{Arv74}, Huttel \etal~\cite{HutArnBau83}, Kievski \etal~\cite{KieRosTor96}, and Wood \etal~\cite{WooBruFis02} for $p$--$d$.}
    \label{fig:Nd32}
\end{figure}

\section{Results: Binding Energies}
Because the two-body LECs are constrained only by low-energy scattering parameters, the deuteron binding energy provides the first genuine prediction of our framework.
We obtain $B_d = 2.213$~MeV, in excellent agreement with the experimental value $B_d = 2.225$~MeV \cite{HuaWanKon21}, differing by less than 1\%.  

As expected in \nopieft, a three-body force (3BF) must be included to correct the overbinding observed in the $A=3,4$ nuclei \cite{KapSavWis98,BedHamKol99}.
We adjust its corresponding LEC $D$ and cutoff $\Lambda_3$ to optimally reproduce the binding energies of the triton ($t \equiv {}^3$H), helion ($h \equiv {}^3$He), and alpha particle ($\alpha \equiv {}^4$He). 
With these parameters fixed, the calculated binding energies are
\[
B_t = 8.421~\text{MeV},\quad 
B_h = 7.685~\text{MeV},\quad 
B_\alpha = 28.423~\text{MeV},
\]
to be compared with the experimental values \cite{HuaWanKon21},
\[
B_t = 8.482~\text{MeV},\quad 
B_h = 7.718~\text{MeV},\quad 
B_\alpha = 28.296~\text{MeV}.
\]
These values deviate from experiment by less than $1\%$ for $A=3$ and at the $0.5\%$ level for $A=4$.

Table~\ref{tab:BE} summarizes the calculated binding energies for $A=2$--4 systems, both with and without the three-body force, alongside experiment.

\begin{table}
    \caption{\label{tab:BE} 
    Binding energies (in MeV) of light nuclei calculated with and without the three-body force (3BF), compared with experiment \cite{HuaWanKon21}.}
    \begin{ruledtabular}
    \begin{tabular}{ccccc}
       System       & $d$    & $t$    & $h$    & $\alpha$ \\ \hline
       Without 3BF  & 2.213  & 9.609  & 8.817  & 36.778   \\
       With 3BF     &  --    & 8.421  & 7.685  & 28.423   \\
       Experiment   & 2.225  & 8.482  & 7.718  & 28.296   \\
    \end{tabular}
    \end{ruledtabular}
\end{table}

The simultaneous reproduction of the triton and alpha-particle binding energies follows the well-known Tjon correlation between $A=3$ and $A=4$ systems~\cite{Tjo75,PlaHamMei05}.

\section{Results: Few-Nucleon Scattering}
Few-nucleon scattering has long provided a sensitive testing ground for nuclear forces and for assessing the applicability of EFTs.
Extensive theoretical and experimental studies of nucleon--nucleon ($N$--$N$), nucleon--deuteron ($N$--$d$), and more complex reactions such as nucleon--triton ($N$--$t$), nucleon--helion ($N$--$h$), and deuteron--deuteron ($d$--$d$) scattering have offered valuable insight into the role of two-, three-, and higher-body components of the nuclear interaction. 

\subsection{Nucleon--Nucleon Scattering}
The nucleon--nucleon system represents the simplest and most extensively studied nuclear scattering problem. A wealth of experimental data exists across a wide energy range, and high-precision phase-shift analyses have been performed by several groups~\cite{StoKloRen93,ArnBriStr07,Mac01}. These provide stringent benchmarks for theoretical approaches, including phenomenological potentials, chiral EFT, and \nopieft. 

In our framework, the two-body low-energy constants (LECs) and the finite cutoffs are fitted to reproduce the singlet and triplet scattering lengths as well as the corresponding effective ranges. The first prediction we present concerns proton--proton ($p$--$p$) scattering.  

\subsubsection{Proton--Proton Scattering}
For proton--proton scattering in the spin--singlet $(S=0,\,I=1)$ channel we obtain
\[
a_{pp} = -7.98 \pm 0.03 \fm, \qquad r_{pp} = 2.69 \pm 0.09 \fm,
\]
Here, and throughout, quoted uncertainties reflect only numerical errors; the estimated theoretical uncertainty of our framework is $\sim 10\%$.

EFT-based analysis~\cite{KongRav00} yielded
\[
a_{pp} = -7.82 \fm, \qquad r_{pp} = 2.83 \fm,
\]
while a prior phase-shift analysis of low-energy $p$--$p$ scattering~\cite{BerCamSan88} reported
\[
a_{pp} = -7.8063 \pm 0.0026 \fm, \qquad r_{pp} = 2.794 \pm 0.014 \fm.
\]

Our results are in very good agreement with both EFT-based analyses and experimental phase-shift results, confirming that at the chosen cutoff no dedicated $p$--$p$ counterterm is required.

Having validated our approach in the two-body sector, we now turn to three-body scattering channels, which provide a more stringent test of finite-cutoff \nopieft.

\subsection{Three-Nucleon Scattering}
The three-nucleon system exhibits a range of nontrivial phenomena, including the Thomas collapse, the emergence of genuine three-body forces, and the Efimov effect.  
Scattering processes in this sector thus provide a stringent testing ground for nuclear interactions and, in particular, for \nopieft.  
Extensive work has been carried out on $N$--$d$ scattering within \nopieft\ at various orders; see, e.g., Refs.~\cite{RupKon03,KonHam11,VanEgoKer14,KonHam14,KonGriHam15,SchBaz23}.  

At low energies, nucleon--deuteron scattering occurs in two $S$-wave channels:  
(i) the doublet channel ($S=1/2$), where three-body force enters already at leading order, and  
(ii) the quartet channel ($S=3/2$), where aligned nucleon spins suppress three-body force at leading order.  

Figure~\ref{fig:Nd32} shows our quartet-channel phase shifts together with available experimental data~\cite{PhiBar69,Arv74,HutArnBau83,KieRosTor96,WooBruFis02}.

A distinctive feature of the $nd$ doublet channel is the presence of a nearby pole in the effective-range function, which prevents convergence of the standard effective-range expansion (ERE).  
In such cases one employs the modified effective-range expansion (MERE) introduced in Ref.~\cite{OerSea67},
\begin{equation}
k \cot \delta = -A + \tfrac{1}{2} B\,k^2 - \frac{C}{1 + D\,k^2} + \dots,
\label{eq:mere}
\end{equation}
where $a = 1/(A+C)$, $r = B$, and the final term encodes the pole at $k^2=-1/D$.  
Figure~\ref{fig:Nd} illustrates the MERE fits for $n$--$d$ and $p$--$d$ scattering in the $S=1/2$ channel, along with available experimental data~\cite{OerBro67,Arv74,HutArnBau83,KieRosTor96,WooBruFis02}.

\begin{figure}[t]
    \centering
    \includegraphics[width=\columnwidth]{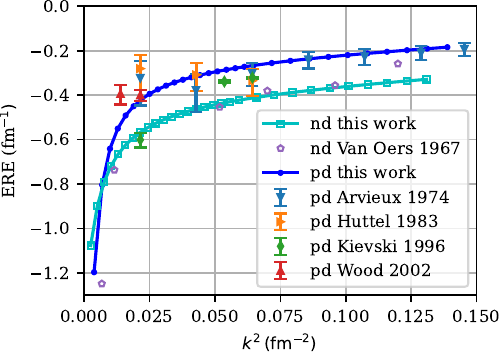}
    \caption{Modified effective-range expansion for nucleon--deuteron scattering in the $S=1/2$ channel. 
    Results for $n$--$d$ (cyan squares) and $p$--$d$ (blue dots) are fitted using Eq.~\eqref{eq:mere} (solid lines). 
    Experimental data are included for comparison: Van~Oers \etal~\cite{OerBro67} ($n$--$d$), and Arvieux~\cite{Arv74}, Huttel \etal~\cite{HutArnBau83}, Kievski \etal~\cite{KieRosTor96}, and Wood \etal~\cite{WooBruFis02} ($p$--$d$).}
    \label{fig:Nd}
\end{figure}

\subsubsection{Neutron--Deuteron Scattering}
For neutron--deuteron scattering in the spin--doublet channel ($S=1/2,\,I=1/2$) we obtain
\[
a_{nd}^{1/2} = 0.71 \pm 0.01 \fm, \qquad r_{nd}^{1/2} = 1.45 \pm 0.02 \fm.
\]
These values are consistent with the uncertainty bands of NLO \nopieft\ calculations~\cite{SchBaz23},
\[
a_{nd}^{1/2} = 0.92 \pm 0.29 \fm, \qquad r_{nd}^{1/2} = 1.74 \pm 0.33 \fm,
\]
and with the experimental determination of Ref.~\cite{Dil71},
\[
a_{nd}^{1/2} = 0.65 \pm 0.04 \fm.
\]
They also agree with the value extracted from the world-average coherent scattering length~\cite{SchJacAri03},
\[
a_{nd}^{1/2} = 0.645 \pm 0.003~(\text{exp}) \pm 0.007~(\text{theory}) \fm.
\]

The $n$--$d$ spin--quartet channel ($S=3/2,\,I=1/2$) is of particular interest due to its connection with universal particle--dimer scattering.  
Following the seminal work of Skorniakov and Ter-Martirosian~\cite{STM}, the zero-range limit of particle--dimer scattering exhibits universal behavior, relating the three-body scattering length $a^{3/2}$ and effective range $r^{3/2}$ to their two-body counterparts in the spin-triplet channel, $a_t$ and $r_t$~\cite{STM,GSS84,Pet03}:
\begin{align}\label{eq:STM}
a^{3/2} &= 1.179066\,a_t - 0.03595\,r_t = 6.326\fm,\\
r^{3/2} &= -0.0383\,a_t + 1.0558\,r_t = 1.643\fm. \nonumber
\end{align}
The above numbers follow from the same $a_t$ and $r_t$ values used to determine the two-body LECs.  

In our calculations, we find
\[
a_{nd}^{3/2} = 6.29 \pm 0.01\fm, \qquad r_{nd}^{3/2} = 2.14 \pm 0.02\fm.
\]
Although the zero-range approximation is not strictly applicable due to the finite-cutoff regularization, the extracted values remain close to the universal predictions of Eq.~\eqref{eq:STM}, especially for the scattering length and, to a lesser degree, for the effective range.  

Our results are also consistent with NLO \nopieft\ predictions~\cite{SchBaz23}:
\[
a_{nd}^{3/2} = 6.322 \pm 0.005\fm, \qquad r_{nd}^{3/2} = 1.875 \pm 0.065\fm,
\]
and with the experimental value~\cite{Dil71},
\[
a_{nd}^{3/2} = 6.35 \pm 0.02\fm.
\]

\subsubsection{Proton--Deuteron Scattering}
For proton--deuteron scattering in the spin--doublet channel ($S=1/2,\,I=1/2$) we obtain
\[
a_{pd}^{1/2} = 0.16 \pm 0.04 \fm, \qquad r_{pd}^{1/2} = 1.30 \pm 0.03 \fm.
\]
These results are compatible with modern analyses of the scattering length and with the well-established value of the effective range~\cite{BlaKarLud99,Arv74},
\[
a_{pd}^{1/2} = -0.13 \pm 0.04 \fm, \qquad r_{pd}^{1/2} = 2.27 \pm 0.12 \fm.
\]
This scattering length also helps resolve earlier discrepancies in the literature~\cite{Arv74,HutArnBau83}.  

For the spin--quartet channel ($S=3/2,\,I=1/2$) we find
\[
a_{pd}^{3/2} = 13.55 \pm 0.01 \fm, \qquad r_{pd}^{3/2} = 1.96 \pm 0.01 \fm.
\]
These values are consistent with NLO \nopieft\ results~\cite{RojSch25},
\[
a_{pd}^{3/2} = 12.76 \pm 0.29 \fm, \qquad r_{pd}^{3/2} = 1.17 \pm 0.07 \fm,
\]
and with experimental determinations~\cite{Arv74,BlaKarLud99},
\[
a_{pd}^{3/2} = 14.7 \pm 2.3 \fm, \qquad r_{pd}^{3/2} = 2.63^{+0.01}_{-0.02} \fm,
\]
as well as with phenomenological models~\cite{ChePayFri91} giving
\[
a_{pd}^{3/2} = 13.76 \pm 0.05\fm, \qquad a_{pd}^{3/2} = 13.52 \pm 0.05\fm.
\]

Taken together, the $n$--$d$ and $p$--$d$ results in both doublet and quartet channels confirm the consistency of our approach with experimental measurements and NLO \nopieft\ predictions, for both neutral and charged three-nucleon scattering.

\subsection{Four-Nucleon Scattering}
Scattering in four-nucleon systems, such as $N$--$t$, $N$--$h$, and $d$--$d$, poses additional challenges due to the increased number of open channels and the complexity of spin--isospin couplings.  
Experimental information is comparatively scarce, but measurements of observables such as total cross sections and analyzing powers provide valuable constraints. 

Theoretical studies of the four-nucleon sector within \nopieft\ have successfully reproduced binding energies and low-energy scattering observables~\cite{PlaHamMei05,Kir13,SchBaz23}.  

As an illustration, Fig.~\ref{fig:4He} shows the spectrum of four nucleons in a harmonic trap with $S=0$, $I=0$, and $L=0$.  
The bound $^4$He state, as well as the scattering states corresponding to $p$+$t$, $d$+$d$, and $d$+$n$+$p$, are clearly visible.

\begin{figure}
    \centering
    \includegraphics[width=\columnwidth]{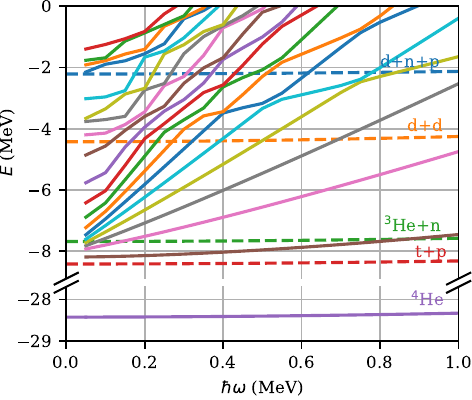}
    \caption{Spectrum of four nucleons in a harmonic trap with $S=0$, $I=0$, and $L=0$.  
    The bound $^4$He state is shown together with the $p$--$t$, $n$--$h$, $d$--$d$, and $d$--$n$--$p$ thresholds (dashed lines), as well as the corresponding trapped energies (solid lines).}
    \label{fig:4He}
\end{figure}

\subsubsection{Neutron--Triton Scattering}
For neutron--triton scattering we obtain in the spin--singlet ($S=0,\,I=1$) channel
\[
a_{nt}^{0} = 4.02 \pm 0.01 \fm, \qquad r_{nt}^{0} = 1.90 \pm 0.01 \fm,
\]
and in the spin--triplet ($S=1,\,I=1$) channel
\[
a_{nt}^{1} = 3.55 \pm 0.01 \fm, \qquad r_{nt}^{1} = 1.62 \pm 0.01 \fm.
\]
These results agree with NLO \nopieft\ predictions~\cite{SchBaz23},
\[
a_{nt}^{0} = 4.035 \pm 0.065 \fm, \qquad r_{nt}^{0} = 2.17 \pm 0.15 \fm,
\]
\[
a_{nt}^{1} = 3.566 \pm 0.047 \fm, \qquad r_{nt}^{1} = 1.76 \pm 0.41 \fm,
\]
and are close to \(\chi\)EFT calculations at \(\text{N}^{3}\text{LO}\)~\cite{VivGirKie20,VivPrivate} without three-body forces,
\[
a_{nt}^{0} = 4.171 \pm 0.063 \fm, \qquad r_{nt}^{0} = 2.117 \pm 0.010 \fm,
\]
\[
a_{nt}^{1} = 3.646 \pm 0.023 \fm, \qquad r_{nt}^{1} = 1.743 \pm 0.010 \fm,
\]
while including a three-body force slightly modifies the values,
\[
a_{nt}^{0} = 4.046 \pm 0.081 \fm, \qquad r_{nt}^{0} = 2.058 \pm 0.004 \fm,
\]
\[
a_{nt}^{1} = 3.533 \pm 0.026 \fm, \qquad r_{nt}^{1} = 1.709 \pm 0.006 \fm.
\]

Experimentally, the most commonly reported observable is the coherent scattering length,
\[
a_{nt}^{c} = \tfrac{1}{4}a_{nt}^{0} + \tfrac{3}{4}a_{nt}^{1},
\]
as well as total cross sections.  
Measurements yield~\cite{HamRauCle81,RauTupWol85}
\[
a_{nt}^{c} = 3.59 \pm 0.02 \fm,
\]
and also provide estimates of the individual channel scattering lengths,
\[
a_{nt}^{0} = 4.98 \pm 0.29 \fm, \qquad a_{nt}^{1} = 3.13 \pm 0.11 \fm.
\]

\subsubsection{Proton--Triton Scattering}
The proton--triton system exhibits a rich channel structure, while the Coulomb interaction introduces additional computational challenges. We calculate the scattering parameters for all spin--isospin channels.

For $S=0,\,I=0$ we obtain
\[
a^{0,0}_{pt} = -31 \pm 2 \fm, \qquad r^{0,0}_{pt} = 2.9 \pm 0.2 \fm,
\]
while for $S=1,\,I=0$ we find
\[
a^{1,0}_{pt} = -4.3 \pm 0.4 \fm, \qquad r^{1,0}_{pt} = -9.0 \pm 0.6 \fm.
\]

In the $S=0,\,I=1$ channel, the effective-range function exhibits a near-threshold pole at $k^2 \approx 0.07 \fm^{-2}$.  
Employing the modified effective-range expansion [Eq.~\eqref{eq:mere}] yields
\[
a^{0,1}_{pt} = -3.6 \pm 0.4 \fm, \qquad r^{0,1}_{pt} = -1.8 \pm 0.8 \fm.
\]
Finally, for $S=1,\,I=1$ we obtain
\[
a^{1,1}_{pt} = -4.8 \pm 0.4 \fm, \qquad r^{1,1}_{pt} = -7 \pm 1 \fm.
\]

To our knowledge, no recent direct measurements of low-energy $p$--$t$ scattering exist.  
Theoretical studies of phase shifts and cross sections have been performed using a variety of approaches~\cite{Laz09,VivDelLaz17,VivGirKie20}.  
For example, Ref.~\cite{Laz09} compared six different potentials, obtaining a spin-triplet scattering length in the range
\[
a_{pt}^{1} \in [5.37, 5.85] \fm,
\]
and a broader spread for the spin-singlet channel,
\[
a_{pt}^{0} \in [-37.4, -15.5] \fm.
\]
Since these results are not projected onto definite isospin, a direct comparison with our values is not straightforward. 

\subsubsection{Neutron--Helion Scattering}
We are unable to extract reliable results for neutron--helion scattering in any spin--isospin configuration (see, for example, Fig.~\ref{fig:4He}), most likely due to the proximity of the $p$--$t$ threshold.

\subsubsection{Proton--Helion Scattering}
For proton--helion scattering we obtain in the spin--singlet ($S=0,\,I=1$) channel
\[
a_{ph}^{0} = 11.1 \pm 0.1\fm, \qquad r_{ph}^{0} = 1.66 \pm 0.03\fm,
\]
and in the spin--triplet ($S=1,\,I=1$) channel
\[
a_{ph}^{1} = 8.9 \pm 0.1\fm, \qquad r_{ph}^{1} = 1.47 \pm 0.02\fm.
\]
These results are in excellent agreement with the NLO \nopieft\ predictions of Ref.~\cite{RojSch25},
\[
a_{ph}^{0} = 11.26 \pm 0.04\fm, \qquad r_{ph}^{0} = 1.65 \pm 0.26\fm,
\]
\[
a_{ph}^{1} = 9.06 \pm 0.04\fm, \qquad r_{ph}^{1} = 1.36 \pm 0.25\fm.
\]

They are also consistent with the available experimental determinations~\cite{FisBruKar06,DanArnCes10},
\[
a_{ph}^{0} = 11.1 \pm 0.5\fm, \qquad r_{ph}^{0} = 1.58 \pm 0.12\fm,
\]
\[
a_{ph}^{1} = 9.04 \pm 0.14\fm, \qquad r_{ph}^{1} = 1.50 \pm 0.06\fm.
\]

\subsubsection{Deuteron--Deuteron Scattering}
For deuteron--deuteron scattering we extract the $S$-wave parameters in both the spin--singlet ($S=0,\,I=0$) and spin--quintet ($S=2,\,I=0$) channels.  

The spin--singlet $d$--$d$ scattering state can be associated with an excited state of the trapped $^4$He spectrum, identifiable by its threshold near $-4.4$~MeV.  
As $d$--$d$ energies approach those of the $p$--$t$ channel, an avoided-crossing pattern emerges, distorting the spectrum in the vicinity of the crossing.  
To reduce these distortions, we discard points near the crossing and retain only energies that lie sufficiently far away.  
The remaining $d$--$d$ states, which are reliably isolated from the $p$--$t$ channel, are then used to extract the scattering parameters.  
Figure~\ref{fig:dd} shows the spectrum of four nucleons in a harmonic trap with $S=0$, $I=0$, and $L=0$, highlighting the deuteron--deuteron and proton--triton scattering channels.

\begin{figure}
    \centering
    \includegraphics[width=\columnwidth]{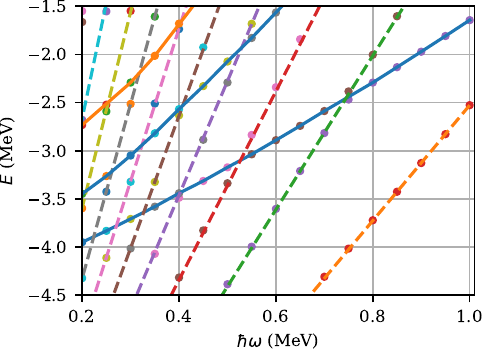}
    \caption{Spectrum of four nucleons in a harmonic trap with $S=0$, $I=0$, and $L=0$.  
    The deuteron--deuteron scattering states are shown by solid lines and the proton--triton states by dashed lines.  
    An avoided crossing is visible when $d$--$d$ energies approach those of the $p$--$t$ channel.}
    \label{fig:dd}
\end{figure}

The extracted spin--singlet parameters are
\[
a_{dd}^{0} = 12.0 \pm 0.5\fm, \qquad r_{dd}^{0} = 2.1 \pm 0.2\fm,
\]
while for the spin--quintet channel we obtain
\[
a_{dd}^{2} = 5.96 \pm 0.02\fm, \qquad r_{dd}^{2} = 1.13 \pm 0.03\fm.
\]

Since, to the best of our knowledge, no direct comparisons are available for the singlet channel, the discussion below focuses on the quintet results.  
These can be directly compared with the NLO \nopieft\ predictions of Ref.~\cite{RojSch25},
\[
a_{dd}^{2} = 6.262 \pm 0.042\fm, \qquad r_{dd}^{2} = 1.41 \pm 0.07\fm.
\]  

Phenomenological approaches have also been pursued. For instance, Ref.~\cite{Filikhin2000} reported
\[
a_{dd}^{2} = 7.5\fm,
\]
while Ref.~\cite{Carew2021} found
\[
a_{dd}^{2} = 7.8 \pm 0.3\fm.
\] 
Earlier studies based on the resonating-group method with the Malfliet--Tjon potential~\cite{MarGenPos75,MeiGlo75} analyzed $d$--$d$ cross sections and extracted both singlet- and quintet-channel $S$-wave phase shifts, finding good agreement with experimental data.  
These phase shifts were reanalyzed in Ref.~\cite{RojSch25}, yielding
\[
a_{dd}^{2} = 6.14 \fm, \qquad r_{dd}^{2} = 1.62 \fm,
\]
in close agreement with our result for $a_{dd}^{2}$, while some deviation is observed for $r_{dd}^{2}$.

\subsection{Five-Nucleon Scattering}
Scattering in five-nucleon systems remains one of the least explored frontiers of low-energy nuclear theory. The simultaneous presence of multiple open channels, strong correlations, and long-range Coulomb effects renders exact treatments of five-nucleon scattering demanding. 
As a result, only limited \textit{ab initio} scattering calculations have been reported to date~\cite{HupLanNav13,HupQuaNav14,ShiMazMaz18,BagSchBaz23,ElhHilMei25,SahMisMoh25}, with most available information derived from phenomenological models.  

In this context, the present study provides systematic investigations of low-energy five-body scattering observables within the \nopieft\ framework. Our results offer valuable benchmarks against experimental data where available and set the stage for future high-precision studies.

\subsubsection{Neutron--Alpha Scattering}
For neutron--alpha scattering in the spin--doublet channel ($S=1/2,\,I=1/2$), we obtain 
\[
a_{n\alpha}^{1/2} = 2.31 \pm 0.01 \fm, \qquad r_{n\alpha}^{1/2} = 1.58 \pm 0.05 \fm,
\]
in good agreement with NLO \nopieft\ results~\cite{BagSchBaz23}:
\begin{align*}
    a_{n\alpha}^{1/2} &= 2.47 \pm 0.04~(\text{num.}) \pm 0.17~(\text{theor.}) \fm, \\
    r_{n\alpha}^{1/2} &= 1.384 \pm 0.003~(\text{num.}) \pm 0.211~(\text{theor.}) \fm. \nonumber
\end{align*}

Recently, a high-precision neutron interferometry experiment~\cite{HauWieAri20} reported
\[
a_{n\alpha}^{1/2} = 2.4746 \pm 0.0017~(\text{stat.}) \pm 0.0011~(\text{syst.}) \fm,
\]
in close agreement with the earlier determination of Ref.~\cite{ArnRop73},
\[
a_{n\alpha}^{1/2} = 2.467 \pm 0.004 \fm,
\]
where the effective range was also extracted,
\[
r_{n\alpha}^{1/2} = 1.28 \pm 0.03 \fm.
\]

Slightly larger values were reported in Ref.~\cite{KaiRauBad79}, which quoted a bound coherent scattering length $b_c = a_{n\alpha}^{1/2}(A+1)/A$:
\[
b_c = 3.26 \pm 0.03 \fm,
\]
corresponding to 
\[
a_{n\alpha}^{1/2} = 2.608 \pm 0.024 \fm.
\]

\subsubsection{Proton--Alpha Scattering}
For proton--alpha scattering in the spin--doublet channel ($S=1/2,\,I=1/2$), we obtain
\[
a_{p\alpha}^{1/2} = 4.43 \pm 0.03 \fm, \qquad r_{p\alpha}^{1/2} = 1.69 \pm 0.06 \fm,
\]
while we are not aware of any other \nopieft\ calculations for comparison.

Experimental scattering parameters were extracted in Ref.~\cite{ArnRopSho71}:
\[
a_{p\alpha}^{1/2} = 4.72 \pm 0.04 \fm, \qquad r_{p\alpha}^{1/2} = 1.36 \pm 0.01 \fm.
\]

Ref.~\cite{KamBay2007} analyzed $p$--$\alpha$ scattering data using the microscopic resonating-group method combined with an $R$-matrix analysis, reproducing earlier results~\cite{BroHaeTra67,SatOweElw68}:
\[
a_{p\alpha}^{1/2} = 4.87 \fm, \qquad r_{p\alpha}^{1/2} = 1.26 \fm,
\]
although no uncertainties were quoted. These values are in good agreement with our calculations.

A summary of our results for the low-energy scattering parameters, alongside available experimental data, NLO \nopieft\ predictions, and other theoretical calculations, is provided in Table~\ref{tab:scatt_summary}.

\begin{table*}[t]
\centering
\caption{Low-energy $S$-wave scattering parameters (in fm) from this work, compared with NLO \nopieft, other theoretical results, and experimental/phase-shift analysis (PSA) data, for systems up to $A=5$.}
\label{tab:scatt_summary}
\begin{tabular}{lccccccc}
\hline\hline
System & Spin & Isospin & Observable & This work & NLO \nopieft\ / Theory & Experiment / PSA \\
\hline
\multicolumn{7}{l}{\textbf{$A=2$}}\\
$p$--$p$ & 0 & 1 & $a_{pp}$ & $-7.98\pm0.03$ & $-7.82$ \cite{KongRav00} & $-7.8063 \pm 0.0026$ \cite{BerCamSan88} \\
         &   &   & $r_{pp}$ & $2.69\pm0.09$  & $2.83$ \cite{KongRav00}  & $2.794\pm0.014$ \cite{BerCamSan88} \\
\hline
\multicolumn{7}{l}{\textbf{$A=3$}}\\
$n$--$d$ & $1/2$ & $1/2$ & $a_{nd}^{1/2}$ & $0.71 \pm0.01$ & $0.92 \pm0.29$  \cite{SchBaz23} & $0.645\pm0.003\pm0.007$ \cite{SchJacAri03} \\
         &       &       & $r_{nd}^{1/2}$ & $1.45 \pm0.02$ & $1.74 \pm0.33$  \cite{SchBaz23} & --- \\
$n$--$d$ & $3/2$ & $1/2$ & $a_{nd}^{3/2}$ & $6.29 \pm0.01$ & $6.322\pm0.005$ \cite{SchBaz23} & $6.35\pm0.02$ \cite{Dil71} \\
        &        &       & $r_{nd}^{3/2}$ & $2.14 \pm0.02$ & $1.875\pm0.065$ \cite{SchBaz23} & --- \\
$p$--$d$ & $1/2$ & $1/2$ & $a_{pd}^{1/2}$ & $0.16 \pm0.04$ & --- & $-0.13\pm0.04$ \cite{BlaKarLud99} \\
        &        &       & $r_{pd}^{1/2}$ & $1.30 \pm0.03$ & --- & $2.27 \pm0.12$ \cite{Arv74} \\
$p$--$d$ & $3/2$ & $1/2$ & $a_{pd}^{3/2}$ & $13.55\pm0.01$ & $12.76\pm0.29$ \cite{RojSch25} & $14.7\pm2.3$ \cite{BlaKarLud99}\\
        &        &       & $r_{pd}^{3/2}$ & $1.96 \pm0.01$ & $1.17 \pm0.07$ \cite{RojSch25} & $2.63^{+0.01}_{-0.02}$ \cite{Arv74}\\
\hline
\multicolumn{7}{l}{\textbf{$A=4$}}\\
$n$--$t$ & 0 & 1 & $a_{nt}^{0}$ & $4.02\pm0.01$ & $4.035\pm0.065$ \cite{SchBaz23}& $4.98\pm0.29$ \cite{RauTupWol85}\\
         &   &   & $r_{nt}^{0}$ & $1.90\pm0.01$ & $2.17 \pm0.15$  \cite{SchBaz23}& --- \\
$n$--$t$ & 1 & 1 & $a_{nt}^{1}$ & $3.55\pm0.01$ & $3.566\pm0.047$ \cite{SchBaz23}& $3.13\pm0.11$ \cite{RauTupWol85}\\
         &   &   & $r_{nt}^{1}$ & $1.62\pm0.01$ & $1.76 \pm0.41$  \cite{SchBaz23}& --- \\
$p$--$t$ & 0 & 0 & $a_{pt}^{0}$ & $-31 \pm2$    & --- & --- \\
         &   &   & $r_{pt}^{0}$ & $2.9 \pm0.2$  & --- & --- \\
$p$--$t$ & 1 & 0 & $a_{pt}^{1}$ & $-4.3\pm0.4$  & --- & --- \\
         &   &   & $r_{pt}^{1}$ & $-9.0\pm0.6$  & --- & --- \\
$p$--$t$ & 0 & 1 & $a_{pt}^{0,1}$ & $-3.6\pm0.4$& --- & --- \\
         &   &   & $r_{pt}^{0,1}$ & $-1.8\pm0.8$& --- & --- \\
$p$--$t$ & 1 & 1 & $a_{pt}^{1,1}$ & $-4.8\pm0.4$& --- & --- \\
         &   &   & $r_{pt}^{1,1}$ & $-7\pm1$    & --- & --- \\
$p$--$h$ & 0 & 1 & $a_{ph}^{0}$ & $11.1\pm0.1$  & $11.26\pm0.04$ \cite{RojSch25} & $11.1\pm0.5$  \cite{DanArnCes10} \\
         &   &   & $r_{ph}^{0}$ & $1.66\pm0.03$ & $1.65 \pm0.26$ \cite{RojSch25} & $1.58\pm0.12$ \cite{DanArnCes10} \\
$p$--$h$ & 1 & 1 & $a_{ph}^{1}$ & $8.9 \pm0.1$  & $9.06 \pm0.04$ \cite{RojSch25} & $9.04\pm0.14$ \cite{DanArnCes10} \\
         &   &   & $r_{ph}^{1}$ & $1.47\pm0.02$ & $1.36 \pm0.25$ \cite{RojSch25} & $1.50\pm0.06$ \cite{DanArnCes10} \\
$d$--$d$ & 0 & 0 & $a_{dd}^{0}$ & $12.0\pm0.5$  & --- & --- \\
         &   &   & $r_{dd}^{0}$ & $2.1 \pm0.2$  & --- & --- \\
$d$--$d$ & 2 & 0 & $a_{dd}^{2}$ & $5.96\pm0.02$ & $6.262\pm0.042$ \cite{RojSch25} & $6.14$ \cite{RojSch25} \\
         &   &   & $r_{dd}^{2}$ & $1.13\pm0.03$ & $1.41 \pm0.07$  \cite{RojSch25} & $1.62$ \cite{RojSch25} \\
\hline
\multicolumn{7}{l}{\textbf{$A=5$}}\\
$n$--$\alpha$ & $1/2$ & $1/2$ & $a^{1/2}_{n\alpha}$ & $2.31\pm0.01$ & $2.47\pm0.04\pm0.17$ \cite{BagSchBaz23} & $2.4746\pm0.0017\pm0.0011$ \cite{HauWieAri20}\\
              &       &       & $r^{1/2}_{n\alpha}$ & $1.58\pm0.05$ & $1.384\pm0.003\pm0.211$ \cite{BagSchBaz23} & $1.28 \pm 0.03 $ \cite{ArnRop73} \\
$p$--$\alpha$ & $1/2$ & $1/2$ & $a^{1/2}_{p\alpha}$ & $4.43\pm0.03$ & --- & 4.87 \cite{KamBay2007} \\
              &       &       & $r^{1/2}_{p\alpha}$ & $1.69\pm0.06$ & --- & 1.26 \cite{KamBay2007} \\
\hline\hline
\end{tabular}
\end{table*}

\subsection{Discussion and Conclusions}
We have implemented the LO \nopieft\ potential with a finite cutoff. 
The two-body LECs and cutoffs are fixed by fitting to the $N$--$N$ scattering lengths and effective ranges in the singlet and triplet channels, while the three-body LEC and cutoff are adjusted to reproduce the binding energies of the triton, helion, and alpha particle.

Using these inputs, we then calculated low-energy $S$-wave scattering parameters for systems with up to five nucleons. The finite-cutoff \nopieft\ framework yields results consistent with the accuracy expected at next-to-leading order in EFT.

In the two-body sector, the proton--proton scattering parameters agree with both experiment and previous EFT analyses, even though no additional $p$--$p$ term is included.  
In the three-body sector, we examined nucleon--deuteron scattering in both spin channels. As expected, no three-body force is required in the $S=3/2$ channel, whereas the $S=1/2$ channel requires a three-body counterterm. In the latter case, the pole near threshold necessitates an anomalous effective-range expansion. Our findings are consistent with earlier studies and available data.  

In the four-body sector, we investigated deuteron--deuteron scattering in the $S=0$ and $S=2$ channels, as well as proton--helion and nucleon--triton scattering. The extracted observables agree with both historical calculations and modern NLO \nopieft\ analyses.  

For the five-body sector, we provided \textit{ab initio} \nopieft\ predictions for nucleon--alpha scattering in the $S=I=1/2$ channel, including $p$--$\alpha$ scattering parameters.  

Overall, these results demonstrate that finite-cutoff \nopieft\ provides a practical and predictive framework for few-body nuclear physics, successfully reproducing binding energies and low-energy scattering observables up to five nucleons. The method combines computational efficiency with systematic improvability, making it a valuable tool for future investigations of light nuclei.  

Looking ahead, natural extensions include applications to larger systems—for instance, testing whether nuclei with $A=6$ and beyond are bound within this framework, as suggested by similar approach in Ref.~\cite{SchGirGne21}. Another promising direction is the description of reactions induced by external probes (e.g., electroweak processes), in the spirit of Ref.~\cite{ParBarCar25}. Systematic studies of higher-order corrections will further enhance the predictive power of finite-cutoff \nopieft, paving the way toward an efficient and quantitatively reliable description of low-energy nuclear dynamics.

\section*{Acknowledgments}
We thank Nir Barnea and Martin Schäfer for useful discussions and comments.  
This work was supported by the Israel Science Foundation (ISF), Grant No.~2441/24.


\end{document}